\documentclass[
    ,final            
  ]
  {aipproc}

\layoutstyle{8x11double}

\begin{document}

\title{Formation Scenario of Magnetars: The Puzzle of Isolation}


\author{Nirvikar Prasad}{
  address={Raman Research Institute, Bangalore 560 080, INDIA}
}

\begin{abstract}
Magnetars (SGRs and AXPs) are one of the most evolutionary paths of a neutron star. These objects have an ultra-strong
magnetic field $B \sim 10^{15}$ G at their surface and show persistent X-ray pulsations and transient bursts. Till date there are 14 magnetars known: 5 SGRs (4 confirmed, 1 candidate) and 9 AXPs (7 confirmed, 2 candidates). It is an open puzzle that all these objects are isolated and none have been found in binaries. We discuss the formation scenario which can lead to such a situation.
\end{abstract}



\maketitle


\section{Introduction}

Magnetars (SGRs and AXPs) are neutron stars having a surface magnetic field, $B \sim 10^{15}$ G. SGRs are high-energy 
transient burst sources and in quiescent phase are X-ray pulsars whereas the AXPs are a quieter version and show only persistent X-ray pulsations \cite{Woods:2006}. The spin periods of magnetars lie in a narrow range of (5 -- 12s). Most of them lie close to the galactic plane implying that magnetars are young objects. All of them are isolated. This fact can be attributed to the small statistics available. But if this is not true and magnetars discovered in the future are also found to be isolated then this becomes one of the most interesting puzzles, answers to which will shed light on their formation scenario. If the progenitor of a magnetar has a binary companion then something must be happening during the formation or the subsequent evolution stage which leads to the disruption of the binary. Here we assume that such is the case and qualitatively speculate on the possible scenarios in which this can occur. We first consider the possibility of binary disruption during the different stages which lead to a supernova explosion and a protomagnetar. Then we discuss a magnetar in a binary and look at its evolution.

\section{Magnetar Formation Scenario}

We begin by exploring the case of disruption of the binary due to the supernova explosion of the magnetar progenitor. 
The successful birth of a neutron star involves a number of stages. It starts with core-collapse of about $1.5 M_{\odot}$ of iron-group elements of radius 
$\sim $ few thousand kms which due to the collapse becomes a neutron rich sphere of radius about 50 kms, a proto-neutron star (PNS). The abrupt halt of the collapse of the inner core due to repulsive nuclear forces generates a shock wave which passes through the outer half of the core. Due to photodisintegration and neutrino losses the shock stalls a few thousand kms from the PNS which meanwhile continues to accrete. In the post bounce phase the PNS deleptonises in the Kelvin-Helmholtz timescale $\tau_{KH} \sim 30s$ emitting $\sim 10^{53} ergs$. The PNS contracts, spins up and a neutron star of radius $10$ km is born. The exact mechanism which leads to the reversal of core-collapse into a supernova explosion is not yet fully understood \cite{WJ:nature:06}. 

The basic question is that what could be happening during the \emph{supernova window}, i.e. different stages which are involved in the supernova explosion, which can lead to disruption? 
It has been suggested that magnetar progenitors are massive \cite{Muno:06}. Assuming this to be the case a magnetar binary will have a large mass ratio, $q \equiv M_{1}/M_{2}$. The simplest case for disruption is when the explosion leads to the loss of more than half of the mass of the system $(M_{1} + M_{2})/2$, then the binary will disrupt, since the orbital velocity will exceed the escape velocity of the companion \cite{Blaauw:61}.

Core-collapse is one of the stages in the \emph{supernova window} where understanding about all the variables involved is yet to happen. 
Specifically, there has been a recent revival of interest in the possible role of rotation and magnetic field in the explosion mechanism of core-collapse supernovae. LeBlanc and Wilson \cite{LW:70} found that when a strong magnetic field   is combined with rotation, the core collapse produces an axial jet.  Recent observational studies of the polarization of supernovae and related issues has lead to the firm conclusion that core collapse supernovae are always asymmetric and frequently bi-polar \cite{WYHW:00}. It is quite possible that progenitor rotation and magnetic field along with the progenitor mass and metallicity play an important part in the core collapse and subsequent explosion. In the process of core-collapse strong toroidal magnetic fields will be generated which can produce magnetocentrifugal jets \cite{WMW:03}.  Also the occurrence of magnetorotational instabilities in the core-collapse phase has been investigated \cite{AWML:03}. 
How does the magnetocentrifugal effect translate for the case of magnetars and what special circumstances lead to their production and accompanying energies and kicks are open questions and areas of current research, see for eg. \cite{SKY:07}. One possibility, for example is that in the core-collapse supernova of a magnetar progenitor, the magnetic field and rotation in conjunction with other variables may end up imparting a high kick velocity to the magnetar invariably which may disrupt the binary. 

Another important interval where huge ($\sim 10^{53} ergs$) amounts of energy is emitted is the Kelvin-Helmholtz timescale $\tau_{KH}$ $\sim$ $GM^{2}/L_{\nu}R$ where $L_{\nu}$ is the total neutrino luminosity. This is where the PNS deleptonises and the compact star is born. In the context of magnetars this is the interval when the strong poloidal magnetic fields are established via the $\alpha$--$\Omega$ dynamo \cite{Duncan:1992}.  
The protomagnetar which is born rapidly rotating lose their rotational energy, $\sim 10^{52} ergs$, efficiently to the co-rotating magnetized wind \cite {TCQ:04}. The spindown time-scale, $\tau _{J}$ $\sim$ $\dot{\omega}/\omega$ $\sim$ (2/5)($\dot{M}/M$)$(r_{NS}/r_A)^{2}$
where $r_{NS}$ and $r_A$ are the neutron star radius and the Alfv\'{e}n radius, is of the order of few seconds. The velocity of this wind is near $c$ and it drives an energetic shock into the slower supernova shock and a hyperenergetic supernova results. If this somehow triggers extreme mass loss then disruption of the binary will occur. Of course a great deal of work needs to be done before one can ascertain the correct channel of magnetar formation and find the possible link to its isolation.

\section{Magnetar Binary Evolution}

Consider the general evolution of a massive binary system. If it is a wide binary then disruption probability is high when the primary explodes in a supernova. On the other hand if it is a close binary then mass transfer will occur from the primary as it overflows its Roche-lobe. The binary evolution is classified as Case A, B, or C depending on the primary state of evolution at the onset of Roche-lobe overflow. The mass transfer will affect the orbital dynamics of the binary and the evolution of the primary and the secondary stars. 

Generally three modes of mass transfer in the evolution of the binary are possible, viz., conservative, quasi-conservative and the common envelope or the non-conservative evolution. Since the components differ significantly in both mass and size a common envelope (CE) evolution at some stage is very likely \cite{TS:ARAA 00}. We consider a magnetar in an evolved binary where we assume that the CE has formed and discuss the possible ways it will evolve. 

Let us first consider a general picture of CE evolution \cite{TS:ARAA 00}. Due to gravitational torques, as the compact star approaches the surface of its larger companion, the orbital separation decreases slowly. We know from the theory of stellar structure that as a star evolves the fraction of stellar matter near the surface increases. Thus more evolved stars can exert greater gravitational torques on the neutron star. The orbital decay timescale, which decreases as the gravitational torques increases, is thus dependent on the evolutionary state of the companion star. For more evolved giant configuration the orbital decay timescale becomes shorter and the spiral-in is faster. Eventually a tightly bound spiral results as the orbital velocities of the two cores exceed the rotational velocity of the envelope. This rapid decay accelerates the gas near the neutron star and the core of the giant which leads to super sonic velocities and the generation of shock. These shocks transfer orbital angular momentum to the spin of the CE and the CE is spun up. 

As a result the matter near the two cores is spun up and this leads to the onset of mass loss from the system. This happens because, due to the spin of matter, the effective gravity decreases which leads to unbalanced pressure gradients. Mass loss rates may reach $\sim$ $1 M_{\odot} yr^{-1}$ \cite{TERMAN:95}. As the mass near the two cores clear up gravitational torques decrease which leads to the increase in the orbital decay timescale. Thus the CE ejection timescale becomes comparable or even shorter than the orbital decay timescale. This give enough time to the CE to eject from the system leaving behind two tightly bound cores. Thus the spiraling in of the two cores and coalescence is prevented. For less evolved stars, due to the presence of large amount of mass near the core the mass loss continues which leads to their merger \cite{TS:ARAA 00}.

Now let's look at a magnetar in a binary system undergoing CE evolution. As the magnetar approaches the giant companion the usual processes as explained above will take place. It will form a tightly bound spiral with the core of the giant. The CE will spin up and mass loss from the system will ensue. As the matter near the two cores clears up somewhat, gravitational torques will decrease and hence the orbital decay time will increase. However since the system is tightly bound now, the magnetic field of the magnetar will thread through the CE and start affecting it. The magnetic fields will exert a braking force on motion of the CE. This will lead to an increase of the ejection timescale of the CE. If this ejection timescale becomes greater than orbital decay timescale, then the two cores will spiral together and coalesce. The outcome is that one ends up with either a magnetar (or neutron star) which survives the merger or a black hole. In any case an isolated object results.

\section{Discussions}
We have tried to discuss the possible ways a magnetar may lose its companion if it originally was in a binary. If further observational evidence accumulates that magnetars are isolated objects then it becomes important to find the mechanism by which this is happening. From the observational point of view it is important to look at the nearby regions of an active magnetar. This may give a clue to the possible channel of disruption or merger. In the parameter space the window which leads to the formation of magnetar instead of a normal neutron star is difficult to quantify. However, the successful working of $\alpha$-$\Omega$ dynamo together with the constraint of disruption which results in isolation, will lead to the tightening of required conditions. We have presented a qualitative discussion which requires a lot of work/simulations before anything concrete can be said about the different possibilities. A detailed analysis will appear elsewhere.


\begin{theacknowledgments}
  We thank A. Deshpande and B. Paul for helpful discussions.
\end{theacknowledgments}


\end{document}